\documentclass{appolb}
\usepackage{epsfig}
\usepackage{graphicx}
\usepackage{dcolumn}   
\usepackage{bm}        
\usepackage{amsmath}
\usepackage{amssymb}   
\usepackage{enumitem}
\usepackage{graphicx,mathrsfs}
\usepackage{nicefrac}
\usepackage{multirow}
\usepackage{color}

\newcommand{\pc}[1]{\ensuremath{\left(#1\right)}}

\begin{document}
\title{The QCD phase diagram in the presence of an external magnetic field:
the role of the inverse magnetic catalysis 
\thanks{Presented at the Summer School and Workshop on High Energy Physics at 
the LHC: New trends in HEP, October 21- November 6 2014, 
Natal, Brazil}
}
\author{
M. Ferreira, P. Costa, C. Provid\^encia
\address{Centro de F\'{\i}sica Computacional, 
Department of Physics, University of Coimbra,
P-3004 – 516 Coimbra, 
Portugal
}
}

\maketitle
\begin{abstract}
The effect of an external magnetic field in QCD phase diagram, namely, in the the 
location of the critical end point (CEP) is investigated. 
Using the 2+1 flavor Nambu–-Jona-Lasinio model with Polyakov loop, it is 
shown that when an external magnetic field is applied its effect on the CEP 
depends on the strength of the coupling.
If the coupling depends on the magnetic field, allowing for inverse magnetic 
catalysis, the CEP moves to lower chemical potentials eventually disappearing,  
and the chiral restoration phase transition is always of first order.  
\end{abstract}

\section{Inverse magnetic catalysis in the PNJL model}

The influence of strong external magnetic fields on the structure of the QCD 
phase diagram is a very important field of research due to its consequences on 
several physical phenomena: the measurements in heavy ion collisions at very 
high energies, the behavior of the first stages of the Universe and 
the understanding of compact astrophysical objects like magnetars.

The inclusion of a magnetic field in the Lagrangian density of the 
Nambu--Jona-Lasinio (NJL) model and of the Polyakov--Nambu--Jona-Lasinio (PNJL) 
model gives rise to the Magnetic Catalysis (MC) effect, i.e., the enhancement 
of the quark condensate due to the magnetic field
\cite{Ferreira:2013tba,Ferreira:2013oda,Ferreira:2014exa}, but fails to account for the 
Inverse Magnetic Catalysis (IMC) found in LQCD calculations 
\cite{baliJHEP2012,bali2012PRD,endrodi2013} where the suppression of the quark 
condensate takes place due to the strong screening effect of the gluon interactions.
In order to overcome this discrepancy, it was proposed, by using the SU(2) NJL model 
\cite{Farias:2014eca} and the SU(3) NJL/PNJL models \cite{Ferreira:2014kpa}, 
that the model coupling, $G_s$, can be seen as proportional to the running coupling, 
$\alpha_s$, and consequently, a decreasing function of the magnetic field strength 
allowing to include the impact of $\alpha_s(eB)$ in both models.
Indeed, the strong screening effect of the gluon interactions in the region of 
low momenta weakens the interaction which is reflected into a decrease of the 
scalar coupling with the intensity of the magnetic field \cite{Miransky:2002rp}.

Since there is no LQCD data available for $\alpha_s(eB)$, by using the NJL model 
we can fit $G_s(eB)$ in order to reproduce the pseudocritical chiral transition 
temperatures, $T_c^\chi(eB)$, obtained in LQCD calculations \cite{baliJHEP2012}. 
The resulting fit function that reproduces the $T_c^\chi(eB)$ is 
\begin{equation}
G_s(\zeta)=G_s^0\pc{\frac{1+a\,\zeta^2+b\,\zeta^3}
{1+c\,\zeta^2+d\,\zeta^4}}\,
\label{eq:fit}
\end{equation}
with $a = 0.0108805$, $b=-1.0133\times10^{-4}$, $c= 0.02228$, and 
$d=1.84558\times10^{-4}$ and where $\zeta=eB/\Lambda_{QCD}^2$. 
We also have used $\Lambda_{QCD}=300$ MeV.

\begin{figure}[t]
\centering
    \includegraphics[width=0.45\linewidth,angle=0]{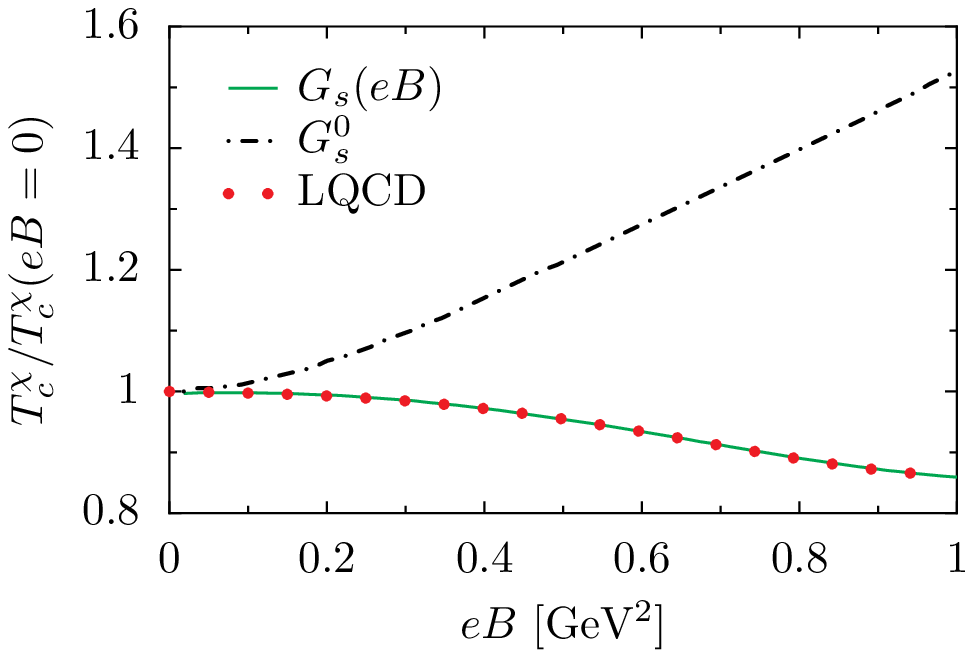}
		\includegraphics[width=0.45\linewidth,angle=0]{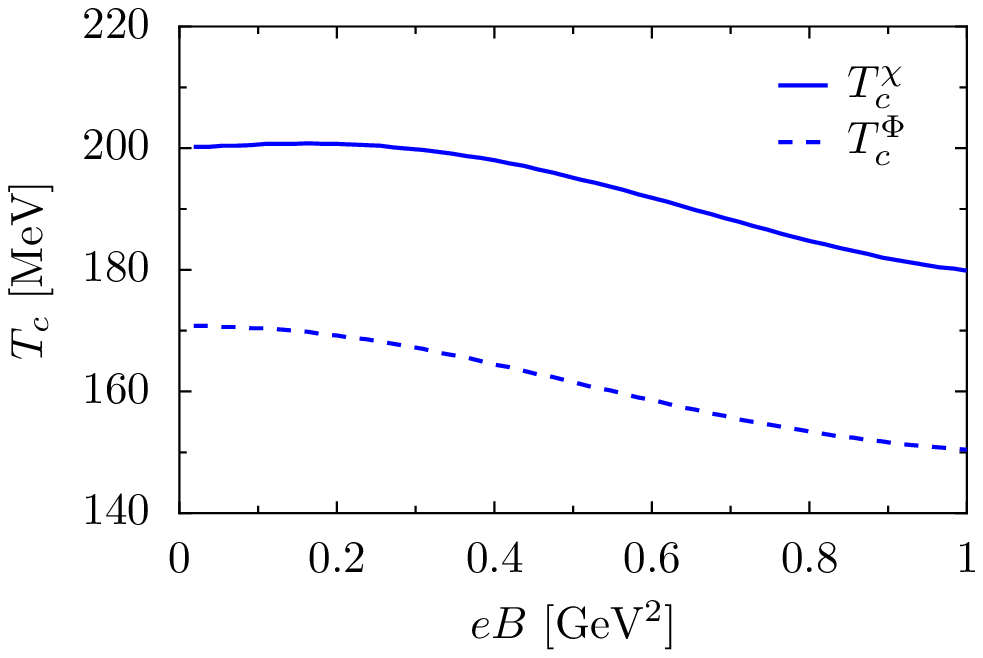}
\caption{
(Left panel) The renormalized critical temperatures of the chiral transition 
($T_c^\chi(eB=0)=178$ MeV) as a function of $eB$ in the NJL model with a
magnetic field dependent coupling $G_s(eB)$ and a constant coupling $G_s^0$, 
and the lattice results \cite{baliJHEP2012}.
(Right panel) The chiral ($T_c^{\chi}$) and deconfinement ($T_c^{\Phi}$) transitions
temperatures as a function of $eB$ in the PNJL, using $G_s(eB)$ given by 
Eq. (\ref{eq:fit}).
}
\label{fig:temp_crit}
\end{figure}

In the NJL model, the renormalized pseudocritical chiral transition temperatures, 
$T_c^{\chi}/T_c^{\chi}(eB=0)$, are plotted in left panel of Fig. \ref{fig:temp_crit}
as a function of $eB$: with the magnetic field dependent coupling $G_s(eB)$ 
(green line), given by Eq. (\ref{eq:fit}); with LQCD results (red dots); and the 
usual constant coupling $G_s=G_s^0$ (black dashed dot line), that shows magnetic 
catalyzes with increasing $T_c^{\chi}/T_c^{\chi}(eB=0)$ for all range of magnetic 
fields. 

Now, using $G_s(eB)$ given in Eq. (\ref{eq:fit}), we calculate the chiral and 
deconfinement transitions temperatures as a function of $eB$ in the PNJL model. 
The results are shown in the right panel of Fig. \ref{fig:temp_crit}: due to the 
existing coupling between the Polyakov loop field and quarks within the PNJL model, 
the $G_s(eB)$ does not only affect the chiral transition but also the deconfinement 
transition. Consequently, both temperatures transitions decrease with increasing 
magnetic filed strength. 

\section{The influence of the inverse magnetic catalysis in the location of the 
critical end point}

The nature of the phase transition and the existence of the critical end point 
(CEP) are open issues for theoretical studies about the QCD phase diagram
\cite{Costa:2015bza}.
From the experimental point of view the existence/location of the CEP is also a 
very timely topic. This renders important to know the conditions 
that can change the position of the CEP in the phase diagram, namely the presence
of strong magnetic fields.

In the following, we will study two scenarios for the effect of a static external 
magnetic field on the location of the CEP when symmetric matter 
($\mu_u=\mu_d=\mu_s$) is considered:\\
Case I $-$ where we take the usual $G_s=G_s^0$ and no IMC effects are included;\\
Case II $-$ where we will use $G_s(eB)$ given by Eq. (\ref{eq:fit}) which will 
allow us to consider the IMC effects on the QCD phase diagram.

\begin{figure*}[tb]
	\centering
  \includegraphics[width=0.46\linewidth,angle=0]{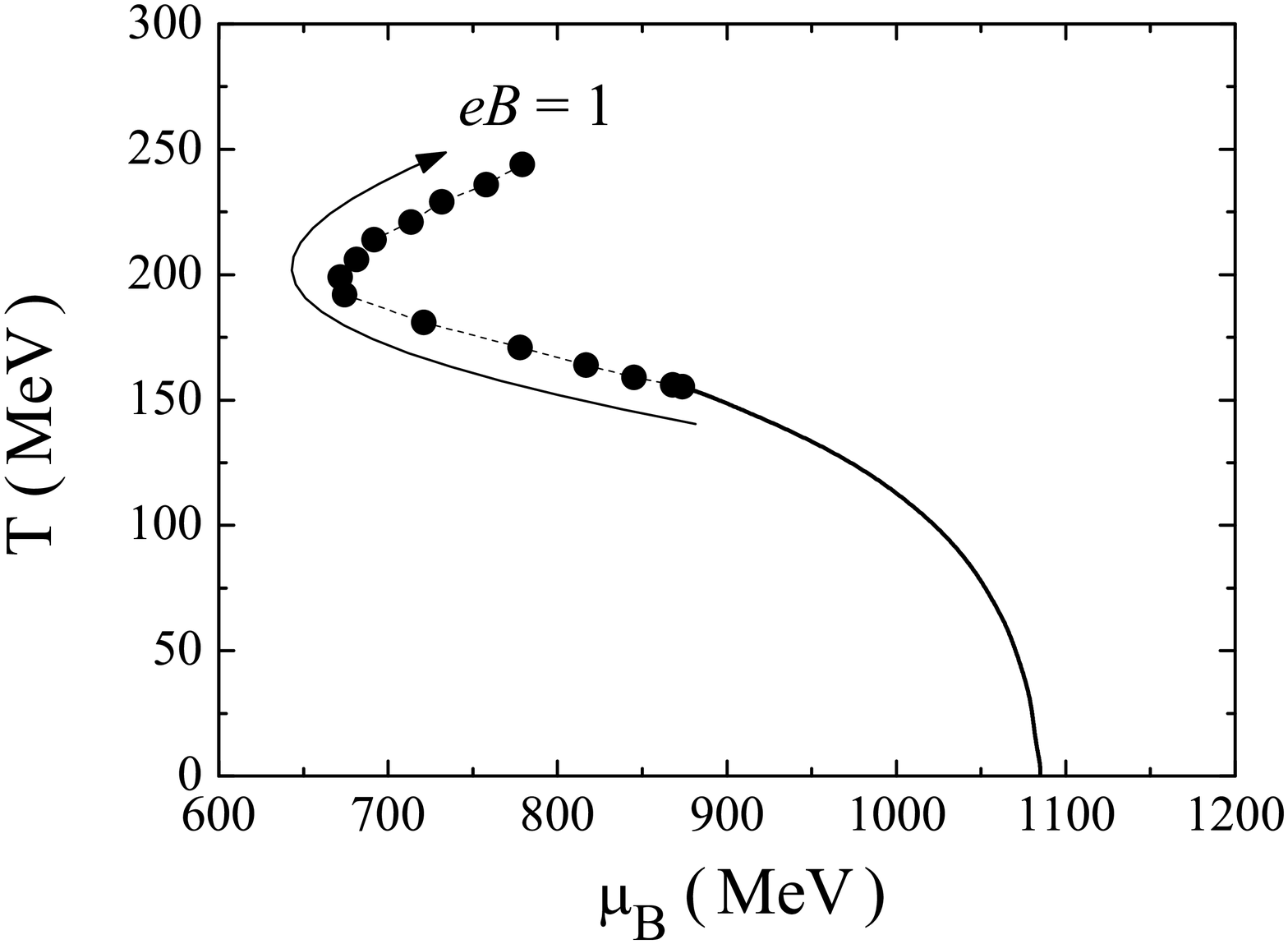} 
	\includegraphics[width=0.46\linewidth,angle=0]{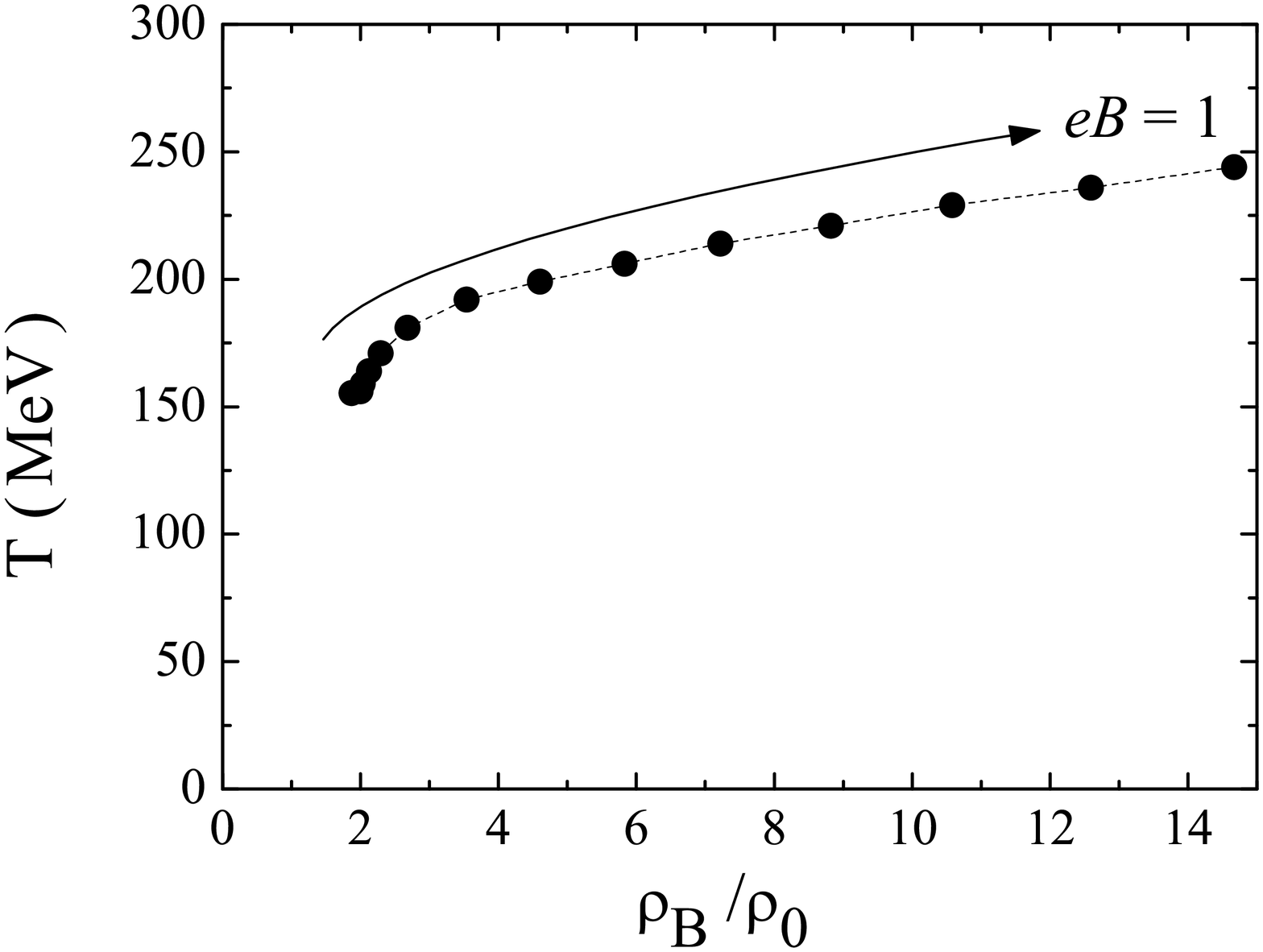}
	\caption{Location of the CEP on temperature vs baryonic
	chemical potential $\mu_B$ (left) and temperature vs 
	baryonic density $\rho_B$ (right) diagrams, for Case I. The 
	baryonic density $\rho_B$ is in units of nuclear saturation density, 
	$\rho_0=0.16$ fm$^{-3}$.} 
	\label{CEP}
\end{figure*}

The results for Case I are plotted in the left panel of Fig. \ref{CEP} and 
reproduce qualitatively the results previously obtained within the NJL model in 
\cite{Avancini:2012ee}: 
as the intensity of the magnetic field  increases, the transition temperature 
increases and the baryonic chemical potential decreases until the critical value 
$eB\sim 0.4$ GeV$^2$. 
For stronger magnetic fields both $T$ and $\mu_B$ increase. 
In the right panel of Fig. \ref{CEP} the CEP is given in a $T$ versus baryonic 
density plot. It is seen that when $eB$ increases from 0 to 1 GeV$^2$ the 
baryonic density at the CEP increases from 2$\rho_0$ to $\sim 14\rho_0$ 
\cite{Costa:2013zca}. 

\begin{figure*}[tb]
	\centering
  \includegraphics[width=0.46\linewidth,angle=0]{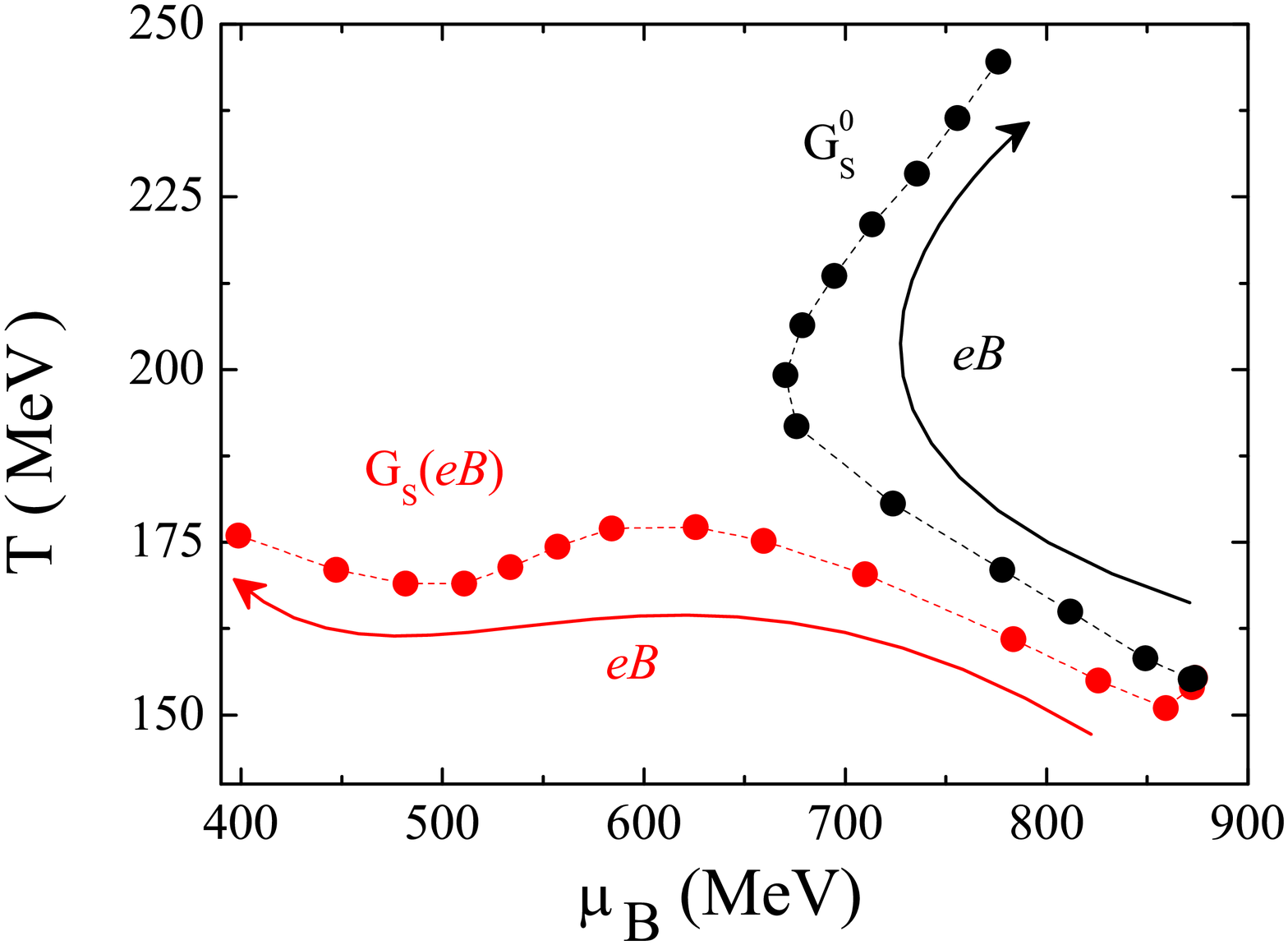} 
	\includegraphics[width=0.46\linewidth,angle=0]{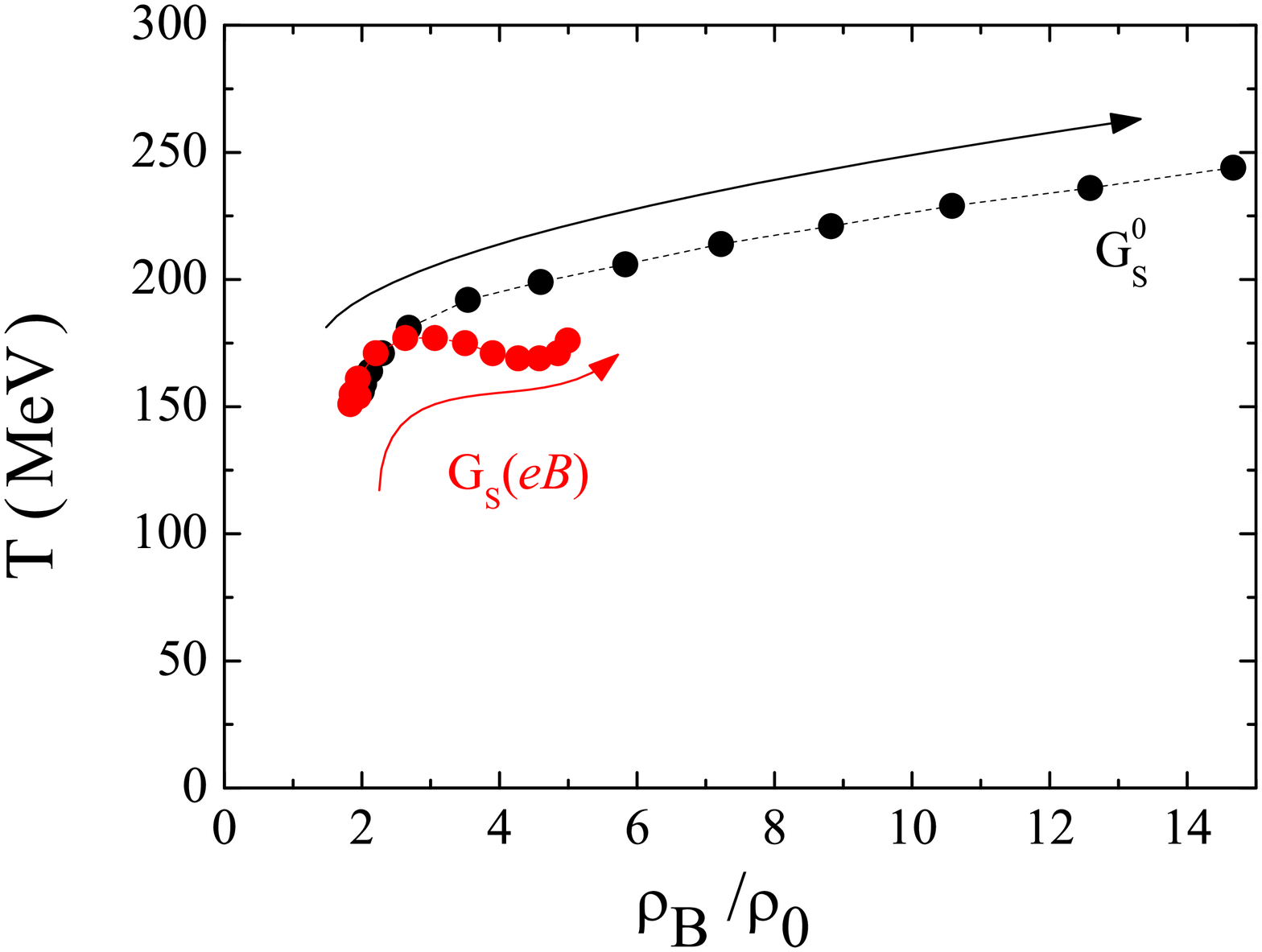} 
	\caption{Location of the CEP on temperature vs baryonic
	chemical potential $\mu_B$ (left) and temperature vs 
	baryonic density $\rho_B$ (right) diagrams, for both cases. The 
	baryonic density $\rho_B$ is in units of nuclear saturation density, 
	$\rho_0=0.16$ fm$^{-3}$.} 
	\label{CEP_IMC}
\end{figure*}

With respect to Case II the results for the CEP are presented in Fig. \ref{CEP_IMC}, 
red points.
We clearly observe a different behavior when compared with Case I (black points): 
at $B=0$ both CEP's coincide but, already for small values of 
$B$, the CEP is moved to lower temperatures and chemical potentials. 
Nevertheless, until $eB\sim 0.3$ GeV$^2$ the pattern is similar for both Cases. 
However, for stronger magnetic fields the position of the CEP in Case II 
oscillates between $T\approx 169$ and $T\approx 177$ MeV while the chemical 
potential takes increasingly smaller values: a completely different behavior 
when compared with Case I, where both values of $T$ and $\mu_B$ for the CEP 
increase.

\begin{figure*}[tb]
	\centering
  \includegraphics[width=0.46\linewidth,angle=0]{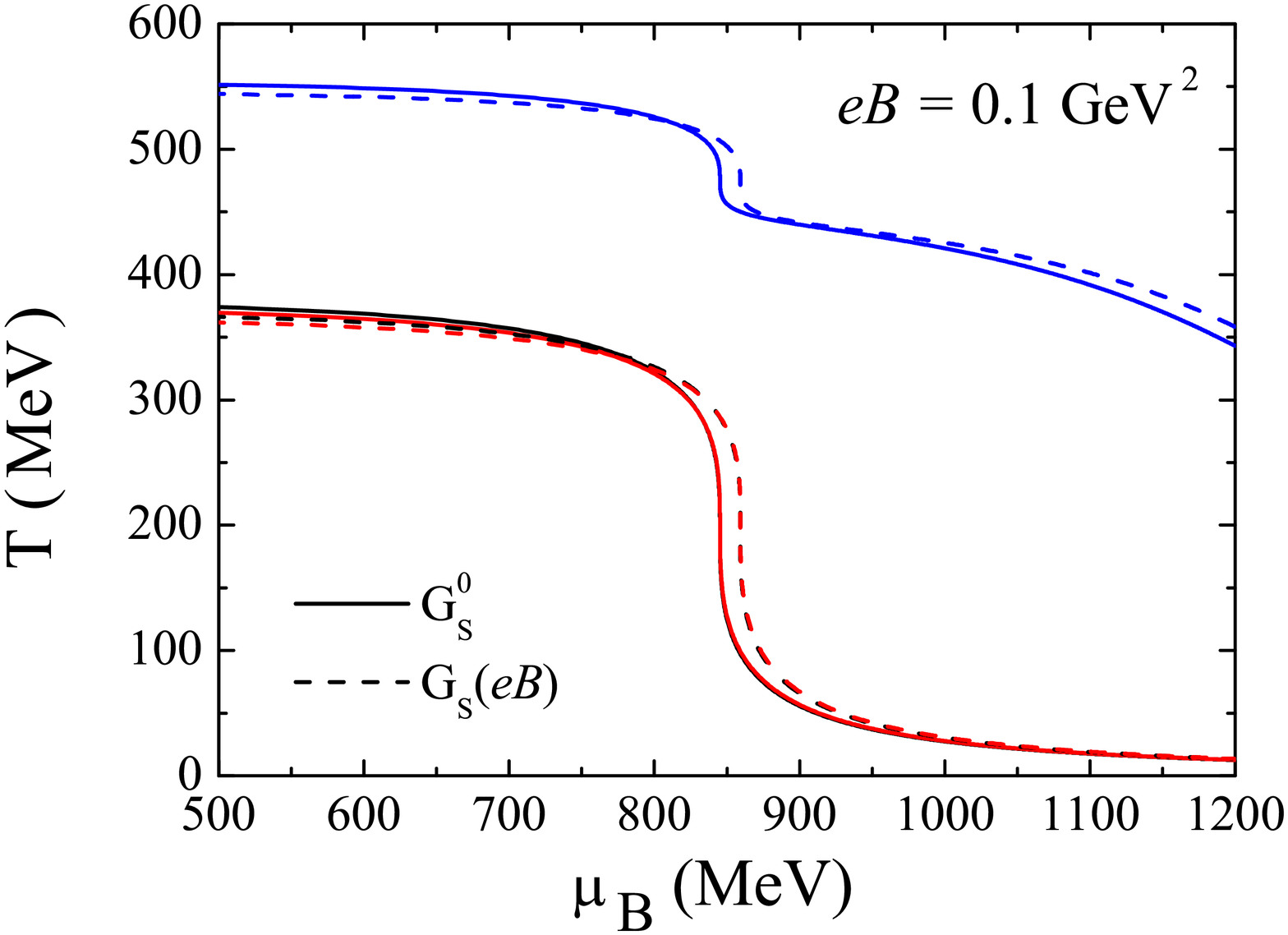} 
	\includegraphics[width=0.46\linewidth,angle=0]{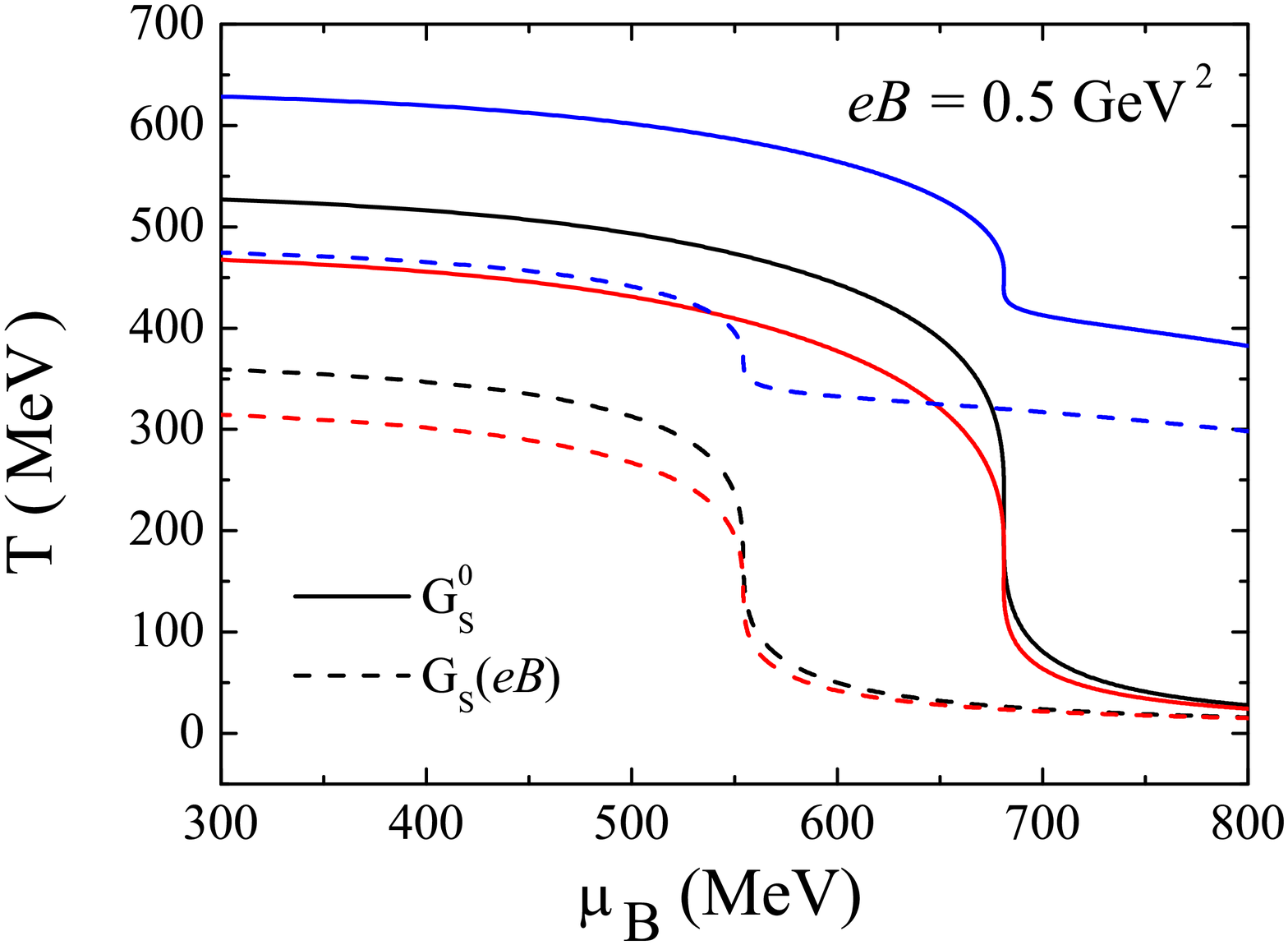} 
	\caption{Masses of the quarks as function of $\mu_B$ for the respective
	$T^{CEP}$ for both Cases.} 
	\label{CEP_mass}
\end{figure*}

The reason of this behavior lies in fact that the restoration of chiral symmetry 
is stressed by the decreasing of the coupling $G_s(eB)$. The increasing of the
magnetic filed is not sufficient to counteract this effect as can be seen if Fig.
\ref{CEP_mass}, where we plot the quarks masses ($M_u$-black line; $M_d$-red line;
$M_s$-blue line) as function of $\mu_B$ for the respective temperature where the
CEP occurs ($T^{CEP}$) at $eB=0.1$ and $eB=0.5$ GeV$^2$. 
At $eB=0.1$ GeV$^2$ (left panel) $G_s$ is barely affected by the magnetic field,
the values of the quark masses are very close to each other for both cases, 
and the CEP occurs at smaller temperatures and at close, but smaller, chemical 
potentials.
When $eB=0.5$ GeV$^2$, the quark masses in Case I are increased with respect to 
the $B=0$ case (due to MC effect), being the restoration of chiral symmetry 
more difficult to achieve.
However, when $G_s(eB)$, Case II, the masses of the quarks are already smaller 
than the $B=0$ case (due to IMC effect) leading to an faster 
restoration of chiral symmetry at small temperatures and chemical potentials. 
Eventually, with the increase of $B$, the CEP would disappear in the temperature 
axis and the transition to the chiral restored phase is always of first order.

\subsection*{Acknowledgement}

This work was partially supported by Project No. PEst-OE/FIS/UI0405
/2014 developed under the initiative QREN financed by the UE/FEDER through 
the program COMPETE $-$ ``Programa Operacional Factores de
Competitividade'', and by Grant No. SFRH/BD/51717/2011 from FCT.


\end{document}